\documentclass[12pt,a4]{article}
\usepackage{epic}
\usepackage{eepic}
\usepackage{epsfig}
\psfigdriver{dvips}
\begin{document}
\renewcommand{\theequation}{\thesection.\arabic{equation}}
\thispagestyle{empty}
\vspace*{-1.5cm}
\hfill {\small KL--TH 98/3} \\[8mm]

\message{reelletc.tex (Version 1.0): Befehle zur Darstellung |R  |N, Aufruf z.B. \string\bbbr}
%
%
\message{reelletc.tex (Version 1.0): Befehle zur Darstellung |R  |N, Aufruf z.B. \string\bbbr}
%
%
%
%
%
\font \smallescriptscriptfont = cmr5
\font \smallescriptfont       = cmr5 at 7pt
\font \smalletextfont         = cmr5 at 10pt
\font \tensans                = cmss10
\font \fivesans               = cmss10 at 5pt
\font \sixsans                = cmss10 at 6pt
\font \sevensans              = cmss10 at 7pt
\font \ninesans               = cmss10 at 9pt
\newfam\sansfam
\textfont\sansfam=\tensans\scriptfont\sansfam=\sevensans
\scriptscriptfont\sansfam=\fivesans
\def\sans{\fam\sansfam\tensans}
\def\bbbr{{\rm I\!R}} 
\def\bbbn{{\rm I\!N}} 
\def\bbbE{{\rm I\!E}} 
\def\bbbm{{\rm I\!M}}
\def\bbbh{{\rm I\!H}}
\def\bbbk{{\rm I\!K}}
\def\bbbd{{\rm I\!D}}
\def\bbbp{{\rm I\!P}}
\def\bbbone{{\mathchoice {\rm 1\mskip-4mu l} {\rm 1\mskip-4mu l}
{\rm 1\mskip-4.5mu l} {\rm 1\mskip-5mu l}}}
\def\bbbc{{\mathchoice {\setbox0=\hbox{$\displaystyle\rm C$}\hbox{\hbox
to0pt{\kern0.4\wd0\vrule height0.9\ht0\hss}\box0}}
{\setbox0=\hbox{$\textstyle\rm C$}\hbox{\hbox
to0pt{\kern0.4\wd0\vrule height0.9\ht0\hss}\box0}}
{\setbox0=\hbox{$\scriptstyle\rm C$}\hbox{\hbox
to0pt{\kern0.4\wd0\vrule height0.9\ht0\hss}\box0}}
{\setbox0=\hbox{$\scriptscriptstyle\rm C$}\hbox{\hbox
to0pt{\kern0.4\wd0\vrule height0.9\ht0\hss}\box0}}}}

\def\bbbe{{\mathchoice {\setbox0=\hbox{\smalletextfont e}\hbox{\raise
0.1\ht0\hbox to0pt{\kern0.4\wd0\vrule width0.3pt height0.7\ht0\hss}\box0}}
{\setbox0=\hbox{\smalletextfont e}\hbox{\raise
0.1\ht0\hbox to0pt{\kern0.4\wd0\vrule width0.3pt height0.7\ht0\hss}\box0}}
{\setbox0=\hbox{\smallescriptfont e}\hbox{\raise
0.1\ht0\hbox to0pt{\kern0.5\wd0\vrule width0.2pt height0.7\ht0\hss}\box0}}
{\setbox0=\hbox{\smallescriptscriptfont e}\hbox{\raise
0.1\ht0\hbox to0pt{\kern0.4\wd0\vrule width0.2pt height0.7\ht0\hss}\box0}}}}

\def\bbbq{{\mathchoice {\setbox0=\hbox{$\displaystyle\rm Q$}\hbox{\raise
0.15\ht0\hbox to0pt{\kern0.4\wd0\vrule height0.8\ht0\hss}\box0}}
{\setbox0=\hbox{$\textstyle\rm Q$}\hbox{\raise
0.15\ht0\hbox to0pt{\kern0.4\wd0\vrule height0.8\ht0\hss}\box0}}
{\setbox0=\hbox{$\scriptstyle\rm Q$}\hbox{\raise
0.15\ht0\hbox to0pt{\kern0.4\wd0\vrule height0.7\ht0\hss}\box0}}
{\setbox0=\hbox{$\scriptscriptstyle\rm Q$}\hbox{\raise
0.15\ht0\hbox to0pt{\kern0.4\wd0\vrule height0.7\ht0\hss}\box0}}}}

\def\bbbt{{\mathchoice {\setbox0=\hbox{$\displaystyle\rm
T$}\hbox{\hbox to0pt{\kern0.3\wd0\vrule height0.9\ht0\hss}\box0}}
{\setbox0=\hbox{$\textstyle\rm T$}\hbox{\hbox
to0pt{\kern0.3\wd0\vrule height0.9\ht0\hss}\box0}}
{\setbox0=\hbox{$\scriptstyle\rm T$}\hbox{\hbox
to0pt{\kern0.3\wd0\vrule height0.9\ht0\hss}\box0}}
{\setbox0=\hbox{$\scriptscriptstyle\rm T$}\hbox{\hbox
to0pt{\kern0.3\wd0\vrule height0.9\ht0\hss}\box0}}}}

\def\bbbs{{\mathchoice
{\setbox0=\hbox{$\displaystyle     \rm S$}\hbox{\raise0.5\ht0\hbox
to0pt{\kern0.35\wd0\vrule height0.45\ht0\hss}\hbox
to0pt{\kern0.55\wd0\vrule height0.5\ht0\hss}\box0}}
{\setbox0=\hbox{$\textstyle        \rm S$}\hbox{\raise0.5\ht0\hbox
to0pt{\kern0.35\wd0\vrule height0.45\ht0\hss}\hbox
to0pt{\kern0.55\wd0\vrule height0.5\ht0\hss}\box0}}
{\setbox0=\hbox{$\scriptstyle      \rm S$}\hbox{\raise0.5\ht0\hbox
to0pt{\kern0.35\wd0\vrule height0.45\ht0\hss}\raise0.05\ht0\hbox
to0pt{\kern0.5\wd0\vrule height0.45\ht0\hss}\box0}}
{\setbox0=\hbox{$\scriptscriptstyle\rm S$}\hbox{\raise0.5\ht0\hbox
to0pt{\kern0.4\wd0\vrule height0.45\ht0\hss}\raise0.05\ht0\hbox
to0pt{\kern0.55\wd0\vrule height0.45\ht0\hss}\box0}}}}

\def\bbbz{{\mathchoice {\hbox{$\sans\textstyle Z\kern-0.5em Z$}}
{\hbox{$\sans\textstyle Z\kern-0.5em Z$}}
{\hbox{$\sans\scriptstyle Z\kern-0.3em Z$}}
{\hbox{$\sans\scriptscriptstyle Z\kern-0.3em Z$}}}}
\setlength{\topmargin}{-1.5cm}
\setlength{\textheight}{22cm}
\begin{center}
{\large\bf Exactly solvable dynamical systems in \\
the neighborhood of the Calogero model}\\
\vspace{0.5cm}
{\large Oliver Haschke and Werner R\"uhl}\\
Department of Physics, University of Kaiserslautern, P.O.Box 3049\\
67653 Kaiserslautern, Germany \\
\vspace{5cm}
\begin{abstract}
The Hamiltonian of the $N$-particle Calogero model can be expressed in terms of generators of
a Lie algebra for a definite class of representations. Maintaining this Lie algebra, its
representations, and the flatness of the Riemannian metric belonging to the second order
differential operator, the set of all possible quadratic Lie algebra forms is investigated.
For $N=3$ and $N=4$ such forms are constructed explicitly and shown to correspond to exactly 
solvable Sutherland models.
The results can be carried over easily to all $N$.
\end{abstract}
\vspace{3cm}
{\it April 1998}
\end{center}
\newpage
\section{Introduction: The Calogero model}
The Calogero model \cite{1} is a quantum mechanical system of $N$ particles moving on a line. It is
completely integrable and possesses both fermionic (totally antisymmetric) and bosonic (totally 
symmetric) solutions. For a review see \cite{2}. In \cite{3} it has been shown that it can be
reduced to a representation theory problem of Lie algebras, so that both eigenvalues and
eigenfunctions can be extracted from finite dimensional representations of these algebras.

The idea that exact solvability and Lie algebraic representation might be connected was
presented in \cite{4}, which dealt with exactly and quasiexactly solvable elementary models. Their
Hamiltonians were rewritten as quadratic polynomials in Lie algebra differential operators.
In this article it was also discussed first whether this procedure could be inverted: Find new
exactly solvable models from Lie algebras of differential operators. Usually this leads to
kinetic energy terms which are Laplace-Beltrami operators on a curved space. So the task arises
to construct quadratic polynomials from the Lie algebras with the constraint that the curvature
tensor vanishes identically on some dynamically accessible domain ("flat quadratic Lie algebraic forms").
In order to maintain the notion of exact integrability, however, the nontrivial but curvature-free 
Laplace-Beltrami operator should be transformed into a standard Laplace operator by an
explicitly known diffeomorphism. This has to be found from nonlinear partial differential
equations. In the neighborhood of the Calogero models for $N = 3$ and $N = 4$ such approach is feasible as 
we shall show in this article. The solutions we give depend on a deformation parameter $s$ ranging over
the full real axis, the case $s = 0$ belonging to the Calogero models themselves.

In the last year two further articles have appeared dealing with the algebraization of
solvable models \cite{5,6}. We hope that these results can be used to simplify the inverse
issue, the construction of new solvable dynamical models.

Let us now outline the algebraization of the standard (sl(N)) Calogero model, since we use
it in the sequel. The Hamiltonian is
\begin{equation}
H = + \frac12 [-\Delta + \omega^2x^2] + \sum_{i<j} \frac{g}{(x_i-x_j)^2}
\label{1.1}
\end{equation}
where $x \in \bbbr_N$ and
\begin{equation}
x^2 = \sum^N_{i=1} x^2_i
\label{1.2}
\end{equation}
$g$ is a coupling constant
\begin{equation}
g = \nu(\nu-1) \ge -\frac14
\label{1.3}
\end{equation}
so that
\[
\nu = \frac12 \pm (g + \frac14)^{\frac12}
\]
is an integer. If $\nu$ is odd (even) we find totally antisymmetric (symmetric) solutions.

We factorize the solutions as
\begin{equation}
\psi(x) = V(x)^{\nu} e^{-\frac12 \omega x^2} P(x)
\label{1.4}
\end{equation}
where $V(x)$ is the Vandermonde determinant
\begin{equation}
V(x) = \prod_{i < j} (x_i-x_j)
\label{1.5}
\end{equation}
and $P(x)$ is a polynomial which is totally symmetric in $\{x_i\}^N_1$. If $\psi$ is a 
solution of
\begin{equation}
H\psi = E\psi
\label{1.6}
\end{equation}
then $P$ has to solve
\begin{equation}
hP = \epsilon P
\label{1.7}
\end{equation}
with
\begin{equation}
h = -\Delta + 2\omega \sum^N_{i=1} x_i \frac{\partial}{\partial x_i} - \nu \sum_{i \not= j}
\frac{1}{x_i-x_j} \left( \frac{\partial}{\partial x_i} - \frac{\partial}{\partial x_j} \right)
\label{1.8}
\end{equation}
and
\begin{equation}
\epsilon = 2E - N\omega - \nu N(N-1)\omega
\label{1.9}
\end{equation}

Next we introduce elementary symmetric functions (following {\cite{3})
\begin{equation}
\prod^N_{i=1} (1 + x_it) = \sum^N_{n=0} \sigma_n(x) t^n \,(\sigma_0=1)
\label{1.10}
\end{equation}
For each element $g \in S_N$ (the symmetric group of $N$ elements) we have a "sector"
\begin{eqnarray}
E_g = \Big\{ x_{i_1} < x_{i_2} < x_{i_3} < \ldots < x_{i_N} ; \nonumber \\
(1,2,3, \ldots N) \begin{array}{c} \\ \to \\ g \end{array} (i_1,i_2,\ldots i_N) \Big\} \nonumber \\
\subset \bbbr_N \quad\quad\quad\quad
\label{1.11}
\end{eqnarray}
so that the map
\begin{equation}
\sigma: x \to \{\sigma_n\}^N_{n=1}
\label{1.12}
\end{equation}
is a diffeomorphism. This follows from the fact that the Jacobian
\begin{equation}
{\cal M}_{ai} = \frac{\partial \sigma_a}{\partial x_i}
\label{1.13}
\end{equation}
has
\begin{equation}
\det {\cal M} = V(x) (-1)^{[\frac N2]}
\label{1.14}
\end{equation}
where $V(x)$ is the Vandermonde determinant. Its square is a polynomial $v(\sigma_1, \ldots,
\sigma_N)$. The domain $D_+ \subset \bbbr_N$ where $v > 0$ is the image $\sigma(E_g)$ for
each $g \in S_N$.

Next we eliminate the centre-of-mass motion by the substitution
\begin{equation}
x_i \to y_i = x_i - \frac 1N \sigma_1
\label{1.15}
\end{equation}
Since
\begin{equation}
\prod^N_{i=1} (1 + y_it) = 1 + \sum^N_{n=2} \tau_n t^n
\label{1.16}
\end{equation}
implies
\begin{equation}
\tau_k = \sigma_k + \sum^{k-1}_{r=0} \left( {N-r \atop k-r} \right) \left( - \frac{\sigma_1}{N}
\right)^{k-r} \sigma_r
\label{1.17}
\end{equation}
we have
\begin{equation}
\det \tilde{\cal M} = V(x) (-1)^{[\frac N2]}
\label{1.18}
\end{equation}
for
\begin{equation}
\tilde{\cal M}_{a,i} = \left\{ \frac{\partial \sigma_1}{\partial x_i} , \frac{\partial \tau_2}
{\partial x_i}, \ldots \frac{\partial \tau_N}{\partial x_i} \right\} 
\label{1.19}
\end{equation}
Therefore we can use
\[ \{\sigma_1, \tau_2, \ldots, \tau_n\} \]
as coordinates on each $E_g$.

On $E_g$ we have the Laplacian (see \cite{3})
\begin{equation}
\Delta = N \frac{\partial^2}{\partial \sigma^2_1} + \sum^N_{j,k=2} A_{jk} \frac{\partial^2}
{\partial\tau_j \partial\tau_k} + \sum^N_{i=2} B_i \frac{\partial}{\partial\tau_i}
\label{1.20}
\end{equation}
with
\begin{eqnarray}
A_{jk} = \frac 1N (N-j+1) (k-1) \tau_{j-1} \tau_{k-1} \nonumber \\
+ \sum_{l \ge \max(1,k-j)} (k-j-2l) \tau_{j+l-1}\tau_{k-l-1} 
\label{1.21}
\end{eqnarray}
\begin{equation}
B_i = - \frac 1N (N-i+2)(N-i+1) \tau_{i-2}
\label{1.22}
\end{equation}
where we insert
\begin{equation}
\tau_0 = 1, \; \tau_i = 0 \; \mbox{for all} \; i \notin \{0,2,3,\ldots,N\}
\label{1.23}
\end{equation}
\newpage
For $h$ (\ref{1.8}) we have then
\begin{eqnarray}
h = - N \frac{\partial^2}{\partial\sigma^2_1} + 2\omega \sigma_1 \frac{\partial}{\partial\sigma_1}
- \sum^N_{j,k=2} A_{jk} \frac{\partial^2}{\partial\tau_j\partial\tau_k} \nonumber \\
+ 2\omega \sum^N_{j=2} j\tau_j \frac{\partial}{\partial\tau_j} + \left( \frac 1N + \nu \right)
\sum^N_{i=2} (N-i+2)(N-i+1)
\tau_{i-2} \frac{\partial}{\partial \tau_i} 
\label{1.24}
\end{eqnarray}

Next we introduce the Lie algebra of operators
\begin{equation}
J_{oi} = \frac{\partial}{\partial\tau_i}, \; J_{ij} = \tau_i \frac{\partial}{\partial\tau_j}, 
\quad i,j \in \{2,3,\ldots, N\}
\label{1.25}
\end{equation}
This algebra has the structure
\begin{equation}
{\it g} l (N-1) \fbox{s} t_{N-1} \quad (t_{N-1} \, {\rm abelian})
\label{1.26}
\end{equation}
We represent it in the polynomial spaces $V_n$
\begin{eqnarray}
V_n &=& \oplus^n_{m=0} U_m \nonumber \\
U_m &=& {\rm span} \left\{ \tau^{n_2}_2, \tau^{n_3}_3, \ldots, \tau^{n_N}_N, \sum^N_{i=2}
n_i = m \right\}
\label{1.27}
\end{eqnarray}
\begin{equation}
\dim V_n = \left( {n + N-1 \atop n }\right)
\label{1.28}
\end{equation}
so that
\[ V_0 \subset V_1 \subset V_2 \subset \ldots \]
forms a flag of polynomial spaces. The algebra (\ref{1.25}) can be obtained from a maximally
degenerate representation of sl(N) (Young tableau with only one row).

Finally we separate off the centre-of-momentum motion by
\begin{equation}
h = - N \frac{\partial^2}{\partial \sigma^2_1} + 2\omega \sigma_1 \frac{\partial}{\partial \sigma_1}
+ h_{\rm rel}
\label{1.29}
\end{equation}
and obtain the relative motion Hamiltonian $h_{\rm rel}$ as a quadratic polynomial in the 
eveloping algebra of (\ref{1.25}), (\ref{1.26}) (see \cite{3}).

\setcounter{equation}{0}
\section{The inverse issue}
We consider a Lie algebra spanned by a basis $\{J_{\alpha}\}^M_1$ being respresented in a
flag of polynomial spaces
\begin{eqnarray}
V_0 \subset V_1 \subset V_2 \subset \ldots \nonumber \\
n_k = \dim V_k < \infty \nonumber \\
1 = n_0 < n_1 < n_2 < \ldots
\label{2.1}
\end{eqnarray}
and such that
\begin{equation}
J^{\alpha} V_n \subset V_n
\label{2.2}
\end{equation}
Then the spectrum and the eigenvectors of the form
\begin{equation}
\sum^M_{\alpha, \beta = 1} C_{\alpha\beta} J_{\alpha} J_{\beta} + \sum^M_{\alpha = 1}
G_{\alpha} J_{\alpha}
\label{2.3}
\end{equation}
with constant coefficients $\{C_{\alpha\beta}, G_{\alpha} \}$ can be calculated from 
representation theory.

The quadratic term in (\ref{2.3}) defines a Laplace-Beltrami operator which we want to
relate with a standard Laplacian from the kinetic energy term in a Schr\"odinger equation. 
For this purpose it must have zero curvature. So the first task is to find the flat quadratic
Lie algebraic forms
\begin{equation}
\sum_{\alpha,\beta} C_{\alpha\beta} J_{\alpha}J_{\beta}
\label{2.4}
\end{equation}
This is a highly non-trivial problem.

We assume that a curvature free quadratic form
\begin{equation}
\sum_{\alpha,\beta} C_{\alpha\beta}^{(o)} J_{\alpha}J_{\beta}
\label{2.5}
\end{equation}
(one point in quadratic Lie algebraic form space) is known. For $t \in \bbbr_M$ define
\begin{equation}
U(t) = \exp \{ \sum_{\alpha} t_{\alpha} J_{\alpha} \}
\label{2.6}
\end{equation}
and its action
\begin{equation}
U(t) J_{\alpha} U(t)^{-1} = \sum_{\beta} T_{\alpha\beta} (t) J_{\beta}
\label{2.7}
\end{equation}
Then
\begin{eqnarray}
U(t)  \left( \sum_{\alpha\beta} C^{(0)}_{\alpha\beta} J_{\alpha}J_{\beta} \right)
U(t)^{-1} \nonumber \\
= \sum_{\alpha\beta} C_{\alpha\beta}(t) J_{\alpha} J_{\beta}
\label{2.8}
\end{eqnarray}
with
\begin{equation}
C_{\alpha\beta}(t) = \sum_{\gamma, \delta} C^{(0)}_{\gamma \delta} T_{\gamma\alpha}(t)
T_{\delta \beta}(t)
\label{2.9}
\end{equation}
defines an $M$-dimensional orbit space of flat quadratic Lie algebraic forms. In the case
(\ref{1.25}), (\ref{1.26})
\begin{equation}
\tau^{\prime}_k = U(t) \tau_k U(t)^{-1} = \Theta_k(t;\tau)
\label{2.10}
\end{equation}
is an affine linear mapping. The flat quadratic Lie algebraic forms we are searching for are
determined up to such Lie algebraic automorphisms.

Assume now that $\{C^{(0)}_{\alpha\beta} \}$ is contained in a curve $\{C_{\alpha\beta}
(s)\}$ of flat quadratic Lie algebraic forms for $s = 0$. Since $\{C^{(0)}_{\alpha\beta} \}$ is
assumed to be flat, there exists a diffeomorphism of some dynamically accessible domain 
\begin{equation}
\tau_i = \phi_i(\xi), \; i \in \{1,2,3,\ldots r\}
\label{2.11}
\end{equation}
so that
\begin{equation}
\sum_{\alpha, \beta} C^{(0)}_{\alpha\beta} J_{\alpha} J_{\beta} = \sum^r_{i=1} 
\frac{\partial^2} {\partial \xi^2_i} + \; \mbox{lower order diff. operator}
\label{2.12}
\end{equation}
Then we need a 1-parametric set of diffeomorphisms
\begin{eqnarray}
\tau_i = f_i(s;\tau^{\prime} ) \nonumber \\
i \in \{1,2,3,\ldots r\}
\label{2.13}
\end{eqnarray}
so that
\begin{equation}
\sum_{\alpha, \beta} C_{\alpha\beta} J_{\alpha}(\tau) J_{\beta}(\tau) = \sum_{\alpha,\beta} 
C^{(0)}_{\alpha\beta} J_{\alpha}(\tau^{\prime}) J_{\beta}(\tau^{\prime}) + \; 
\mbox{lower order diff. operator}
\label{2.14}
\end{equation}
Both types of diffeomorphisms (\ref{2.11}), (\ref{2.13}) ought to be known explicitly, 
in order to maintain the concept of "exact solvability".

In the following sections we shall give a curve of forms for the Calogero point 
$\{C^{(0)}_{\alpha\beta} \}$ for $N = 3$ and $N = 4$ satisfying these requirements.

\setcounter{equation}{0}
\section{The $N=3$ Calogero model and the flatness constraint}
In the case $N=3$ we obtain the following Lie algebraic operators from (\ref{1.25}) in 
lexicographic order
\begin{eqnarray}
J_1 = \frac{\partial}{\partial\tau_2}, \, J_2 = \frac{\partial}{\partial\tau_3}, \, 
J_3 = \tau_2
\frac{\partial}{\partial\tau_2} \nonumber \\
J_4 = \tau_2 \frac{\partial}{\partial\tau_3}, \, J_5 = \tau_3 \frac{\partial}{\partial\tau_2}, 
\, J_6 = \tau_3 \frac{\partial}{\partial\tau_3}
\label{3.1}
\end{eqnarray}
and from (\ref{1.20}), (\ref{1.21}) (then (\ref{2.4}) belongs to $-h$ (\ref{1.8}) and 
$+\Delta$ (\ref{1.20}))
\begin{equation}
C^{(0)}_{16} + C^{(0)}_{25} = -3
\label{3.2}
\end{equation}
\begin{equation}
C^{(0)}_{44}  = + \frac23
\label{3.3}
\end{equation}
\begin{equation}
C^{(0)}_{13}  = -1
\label{3.4}
\end{equation}
all other coefficients are vanishing
\begin{equation}
C^{(0)}_{14} + C^{(0)}_{23} = C^{(0)}_{36} + C^{(0)}_{45} = 0
\label{3.5}
\end{equation}
\begin{eqnarray}
C^{(0)}_{11} = C^{(0)}_{12} = C^{(0)}_{15} = C^{(0)}_{22} = C^{(0)}_{24} = C^{(0)}_{26} =
\nonumber \\
C^{(0)}_{33} = C^{(0)}_{34} = C^{(0)}_{35} = C^{(0)}_{46} = C^{(0)}_{55} = C^{(0)}_{56} = 
C^{(0)}_{66} = 0
\label{3.6}
\end{eqnarray}

In order to find the set $\{C_{\alpha\beta}(s)\}$ corresponding to a flat quadratic Lie algebraic form 
we can calculate the curvature tensor, each component of which is a rational function of the 
$\{C_{\alpha\beta}(s)\}$. Equating a numerator to zero means that in turn all coefficients of 
the monomials $\tau^n_2\tau^m_3$ vanish. This leads to an enormous number of equations of high 
polynomial degree in the 21 variables $\{C_{\alpha\beta}(s)\}$. However, after we discovered 
a solution this way, we developed a simpler algorithm whose variable set is considerably reduced.

First we take account of the irrelevant automorphisms (\ref{2.10}) and postulate that 
(\ref{2.13}) takes the form
\begin{equation}
\tau_i = \tau_i^{\prime} + O(s) \quad i \in \{2, \ldots, N\}
\label{3.7}
\end{equation}
in general. Second we make an ansatz
\begin{eqnarray}
\sum_{\alpha,\beta} C_{\alpha\beta} (s) J_{\alpha} J_{\beta} = \sum^N_{a,b=2} g^{-1}
_{ab} (s;\tau) \frac{\partial}{\partial \tau_a} \frac{\partial}{\partial\tau_b}\nonumber \\
+ \mbox{lower order differential operator}
\label{3.8}
\end{eqnarray}
where each $g^{-1}_{ab}(s;\tau)$ is at most of degree two in $\{\tau_n\}$.
\begin{equation}
g^{-1}_{ab} (s;\tau) = \sum_{i,j\in \{0,2,3,\ldots,N\}} \gamma_{ab,ij} (s) \tau_i\tau_j
\label{3.9}
\end{equation}
\[ (\tau_0 = 1) \]
For $s = 0$ we postulate
\begin{equation}
g^{-1}_{ab}(0;\tau) = (g^{(0)})^{-1}_{ab}(\tau)
\label{3.10}
\end{equation}
and in particular for $N=3$ from (\ref{3.2})-(\ref{3.6})
\begin{equation}
(g^{(0)})^{-1}_{ab} = \left( \begin{array}{cc}
-2 \tau_2 & -3 \tau_3 \\
-3 \tau_3 &  \frac23 \tau^2_2 \end{array} \right)
\label{3.11}
\end{equation}
Third we introduce the concepts of a "dimension" $\lambda_n$ for each $\tau_n$
\begin{equation}
[\tau_n] = \lambda_n
\label{3.12}
\end{equation}
and of a "signature"
\begin{equation}
{\rm sign} (\tau_{2n}) = +1, \quad {\rm sign} (\tau_{2n+1}) = -1
\label{3.13}
\end{equation}
Then the whole expression (\ref{3.8}) is (from (\ref{3.11})) required to be of dimension 
$\lambda^{-1}_2$ and of sign $+1$. The existence of independent dimensions (\ref{3.12}) for each $\tau_n$ is connected with the
fact that (\ref{2.6}), (\ref{2.10}) permits separate dilation automorphisms
\begin{equation}
U_n(t) = \exp \left\{ t \tau_n \frac{\partial}{\partial \tau_n} \right \}
\label{3.14}
\end{equation}
\begin{equation}
\tau^{\prime}_n = e^t\tau_n, \tau^{\prime}_m = \tau_m (m \not= n)
\label{3.15}
\end{equation}

The parameter $s$ has naturally a dimension (see below)
\begin{equation}
s: [s] = \lambda^{-1}_2
\label{3.16}
\end{equation}
But besides $s$ we need other dimensional quantities for $n \in \{3, \ldots,N\}$
\begin{equation}
w_n : [w_n] = \frac{\lambda_2^{\frac n2}}{\lambda_n}
\label{3.17}
\end{equation}
The set
\[ \{s,w_n\}^N_{n=3} \]
contains the "scaling parameters". At the end it will turn out that the diffeomorphism 
(\ref{2.13}) can solely be expressed by functions of scale-invariant variables
\begin{eqnarray}
\xi &=& s \tau_2 \nonumber \\
\eta_n &=& w_n s^{\frac n2} \tau_n, \quad n \in \{3, \ldots,N\}
\label{3.18}
\end{eqnarray}

Using these concepts the most general ansatz for (\ref{3.9}) is for $N = 3$
\begin{equation}
g^{-1}_{ab} (s;\tau) = -\left( \begin{array}{l}
2\tau_2 + s\tau^2_2 + a_2w^2_3s^2\tau^2_3, \, 3 \tau_3 + a_1s\tau_2\tau_3 \\
3\tau_3 + a_1s\tau_2\tau_3, \, - \frac23 w^{-2}_3\tau^2_2 + a_3s\tau^2_3
\end{array} \right)
\label{3.19}
\end{equation}
Obviously neither $s$ nor $w_3$ can be determined from the flatness constraints. In
order to determine the three numerical parameters $a_1, a_2, a_3$ we may set
\begin{equation}
s = 1, \; w_3 = 1
\label{3.20}
\end{equation}
respectively we must obtain (\ref{3.11}) for
\begin{equation}
s = 0, \; w_3 = 1
\label{3.21}
\end{equation}
Then it results from flatness
\begin{equation}
a_1 = \frac43, \; a_2 = - \frac16, \; a_3 = 1
\label{3.22}
\end{equation}
  
\setcounter{equation}{0}
\section{The basic diffeomorphism}
The basic diffeomorphism (\ref{2.13}) is obtained from the system of nonlinear differential
equations
\begin{equation}
g^{-1}(s;\tau) = \Omega^T(g^{(0)})^{-1}(\tau^{\prime})\Omega
\label{4.1}
\end{equation}
where by derivation of (\ref{2.13})
\begin{equation}
\Omega_{ab}(\tau^{\prime}) = \frac{\partial\tau_b}{\partial\tau^{\prime}_a}(\tau^{\prime})
\label{4.2}
\end{equation}
We use the scale invariant variables (\ref{3.18}) with
\begin{equation}
\eta_3 = \eta
\label{4.3}
\end{equation}
Then the Laplace-Beltrami operators appearing in (\ref{2.14}) are
\begin{equation}
\left( \frac{\partial}{\partial\xi}, \frac{\partial}{\partial\eta} \right) \Gamma(\xi,\eta)
\left( {\frac{\partial}{\partial\xi} \atop \frac{\partial}{\partial\eta}} \right)
\quad \mbox{(no differentiation of} \, \Gamma)
\label{4.4}
\end{equation}
with
\begin{equation}
\Gamma(\xi,\eta) = \left( \begin{array}{cc}
-2\xi - \xi^2 + \frac16 \eta^2, & -3\eta - \frac43 \xi \eta \\
-3\eta - \frac43 \xi\eta, & + \frac23 \xi^2 - \eta^2
\end{array} \right)
\label{4.5}
\end{equation}
\begin{equation}
\Gamma^{(0)}(\xi,\eta) = \left( \begin{array} {cc}
-2\xi, & -3\eta \\
-3\eta , & + \frac23 \xi^2
\end{array} \right)
\label{4.6}
\end{equation}
The diffeomorphism (\ref{2.13}) involves two functions
\begin{eqnarray}
\xi &=& F(\xi^{\prime},\eta^{\prime}) \nonumber \\
\eta &=& G(\xi^{\prime}, \eta^{\prime})
\label{4.7}
\end{eqnarray}
that must be determined from
\begin{equation}
\Gamma(\xi,\eta) = H^T \Gamma^{(0)} (\xi^{\prime},\eta^{\prime}) H
\label{4.8}
\end{equation}
where
\begin{equation}
H = \left( \begin{array}{cc}
\frac{\partial F}{\partial \xi^{\prime}} & \frac{\partial G}{\partial \xi^{\prime}} \\
\frac{\partial F}{\partial \eta^{\prime}} & \frac{\partial G}{\partial \eta^{\prime}}
\end{array} \right)
\label{4.9}
\end{equation}
Imposing the boundary conditions (\ref{3.7}) the solution of (\ref{4.8}) is unique and
can be given in series form
\begin{equation}
F(\xi,\eta) = \sum^{\infty}_{\begin{array}{c} {\scriptstyle
n=0} \\ {\scriptstyle (n+m \ge 1)} \end{array}}
 \sum^{\infty}_{m=0} (-1)^m \frac{2^{n+3m-1}(n+2m-1)!}
{n!(2m)!(2n+6m-1)!} \xi^n \eta^{2m} 
\label{4.10}
\end{equation}
\begin{equation}
G(\xi,\eta) = \sum^{\infty}_{n=0} \sum^{\infty}_{m=0} (-1)^m \frac{2^{n+3m+1}(n+2m)!}
{n!(2m+1)!(2n+6m+2)!} \xi^n \eta^{2m+1}
\label{4.11}
\end{equation}
Both functions are entire analytic. The Jacobian $H$ (\ref{4.9}) is, however, not throughout
invertible on the whole $\xi,\eta$-plane.

A summed up form for $F$ and $G$ is obtained after insertion of the functions (see (\ref{1.16}),(\ref{3.18}))
\begin{eqnarray}
\xi & = & s(y_1y_2+y_2y_3+y_3y_1) \label{4.12a} \\ 
\eta & = & s^{\frac{3}{2}}y_1y_2y_3 \label{4.13a} 
\end{eqnarray}
namely
\begin{eqnarray}
F & = & 2 \sum_{\textrm{\tiny{cycl. perm. of }} 1,2,3} \sin{ \{ \left (\frac{s}{2} \right)^\frac12 y_i \} }  \sin{ \{ \left (\frac{s}{2} \right)^\frac12 y_j \} }  \cos{ \{ \left (\frac{s}{2} \right)^\frac12 y_k \} } \label{4.14a} \\
G & = & 8^\frac12 \sin{ \{ \left (\frac{s}{2} \right)^\frac12 y_1 \} } \sin{ \{ \left (\frac{s}{2} \right)^\frac12 y_2 \} } \sin{ \{ \left (\frac{s}{2} \right)^\frac12 y_3 \} } \label{4.15a} 
\end{eqnarray} 

From (\ref{4.10}), (\ref{4.11}) one can easily get the series expansion for $\det H$
\begin{equation}
\det H(\xi,\eta) = \sum^{\infty}_{n,m=0} h_{nm} \xi^n \eta^{2m}
\label{4.12}
\end{equation}
\[ (h_{00} = 1) \]
but a closed form for $h_{nm}$ has not been obtained. Nevertheless we were able to sum this
series as follows.

The square of the Vandermonde determinant gives (at $w_3 = 1$)
\begin{equation}
D = 4\xi^3 + 27\eta^2 = - s^3V(y)^2
\label{4.13}
\end{equation}
so that $s > 0 \, (s < 0)$ implies $D < 0 \, (D > 0)$ respectively. The separatrix
\[ D = 0 \]
separates a domain
\begin{enumerate}
\item[$D < 0$:] here $\det H = 0$ can be fulfilled on an infinite number of zero-trajectories
${\cal C}_n, \, n \in \{1,2,3\ldots \}$;
\end{enumerate}
from the domain
\begin{enumerate}
\item[$D > 0$:] where $\det H > 0$.
\end{enumerate}
The curves ${\cal C}_n$ are given by
\begin{equation}
{\cal C}_n : \left( 1 + \frac32 \frac{\xi}{n^2\pi^2} \right)^2 +
\frac{D}{8n^6\pi^6} = 0
\label{4.14}
\end{equation}
They were first discovered by numerical computations (Fig. 1).
\begin{figure}
\begin{center}
\epsfig{file=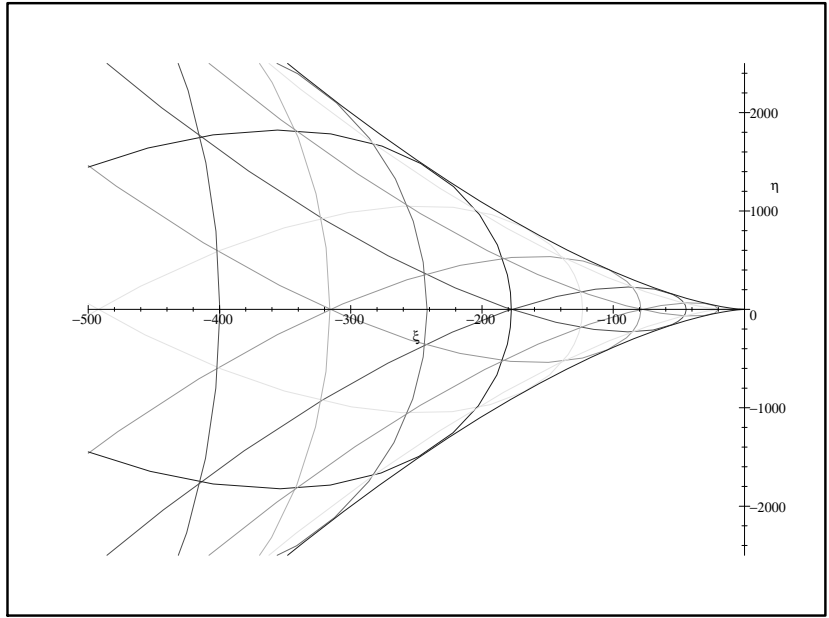,height=12cm,angle=270}
\end{center} 
\label{fig:testbi01}
\caption{The zerotrajectories ${\cal C}_n$ and the separatrix in $(\xi,\eta)$-plane} 
\end{figure}

It is obviously suggesting to attempt a Weierstra{\ss} product ansatz
\begin{equation}
\det H(\xi,\eta) = \prod^{\infty}_{n=1} \left[ \left( 1 + \frac32 \frac{\xi}{n^2\pi^2}
\right)^2 + \frac{D}{8n^6\pi^6} \right]
\label{4.15}
\end{equation}
That this is in fact correct can be verified in the following fashion.

Consider the polynomial
\begin{equation}
x^3 + ux + v = 0
\label{4.16}
\end{equation}
with its factorized form
\begin{equation}
\prod^3_{i=1} (x-\alpha_i) = 0
\label{4.17}
\end{equation}
so the elementary symmetric polynomials of $\{\alpha\}^3_{i=1}$ are
\begin{equation}
\sigma_1(\alpha) = 0, \, \sigma_2(\alpha) = u, \, \sigma_3(\alpha) = -v
\label{4.18}
\end{equation}
By symmetric evaluation of the product we have
\begin{eqnarray}
\prod^{+\infty}_{\begin{array}{l}
{\scriptstyle n=-\infty} \\ {\scriptstyle ({\rm symm.}} \\ {\scriptstyle n \not= 0)} \end{array}}
\left(1 - \frac{\alpha_i}{n\pi} \right) &=& \prod^{\infty}_{n=1} \left( 1 - \frac{\alpha^2_i}
{n^2\pi^2} \right) \nonumber \\
&=& \frac{\sin \alpha_i}{\alpha_i}
\label{4.19}
\end{eqnarray}
Correspondingly we get
\begin{eqnarray}
\det H(\xi,\eta) &=& \prod^{+\infty}_{\begin{array}{l}
{\scriptstyle n=-\infty} \\ {\scriptstyle ({\rm symm.}} \\ {\scriptstyle n \not= 0)} \end{array}}
\left[1 + \frac32 \frac{\xi}{n^2\pi^2} + \sqrt{\frac{-D}{8}} \frac{1}{n^3\pi^3} \right] \nonumber \\
&=& \prod^3_{i=1} \frac{\sin \alpha_i}{\alpha_i}
\label{4.20}
\end{eqnarray}
where 
\begin{equation}
u = \frac32 \xi, \, v = \sqrt{\frac{-D}{8}}
\label{4.21}
\end{equation}
We can expand (\ref{4.20}) into a power series of $\alpha_1, \alpha_2, \alpha_3$. Symmetry makes 
sure that this series can be reformulated in terms of
\[ \sigma_1(\alpha), \, \sigma_2(\alpha), \, \sigma_3(\alpha). \]
Using then (\ref{4.18}) with (\ref{4.21}) we obtain a series expansion as in (\ref{4.12}). By a
numerical comparison we can prove their identity.

Finally we consider the separatrix
\begin{equation}
D = 0 \quad \mbox{implying (say)} \; \alpha_3 = 0, \, \alpha_2 = - \alpha_1
\label{4.22}
\end{equation}
This yields  by (\ref{4.18}), (\ref{4.21})
\begin{equation}
- \sigma_2(\alpha) = \alpha^2_1 = - \frac32 \xi
\label{4.23}
\end{equation}
Then
\begin{equation}
\det H(\xi,\eta)|_{D = 0} = \left( \frac{\sin \alpha_1}{\alpha_1} \right)^2
\label{4.24}
\end{equation}
implying zeros of $\det H$ at $D = 0$ and
\begin{equation}
- \frac32 \xi = (n\pi)^2, \; n \in \bbbz - \{0\}
\label{4.25}
\end{equation}
At this point the zero trajectory ${\cal C}_n$ touches the separatrix.

The representation of the function $\det H(\xi,\eta)$ simplifies if we return to the relative
coordinate variables $y_1,y_2,y_3$. The cubic equation (\ref{4.16}) suggests the
identification
\begin{equation}
\alpha_i = \frac{1}{\sqrt{2}} s^{\frac12} (y_j-y_k)
\label{4.26}
\end{equation}
\[ (i,j,k \; {\rm cyclic}) \]
Then we find easily
\begin{eqnarray}
\sigma_1(\alpha) &=& 0 \nonumber \\
\sigma_2(\alpha) &=& \frac32 s \sigma_2(y) = \frac32 \xi \nonumber \\
\sigma_3(\alpha) &=& - s^{\frac32} \frac{V(y)}{\sqrt{8}} = \mp \sqrt{\frac{-D}{8}}
\label{4.27}
\end{eqnarray}
so that (\ref{4.18}), (\ref{4.21}) are satisfied. The Vandermonde determinant $V(y)$ was 
defined in (\ref{1.5}). In $y$-space the separatrix consists of three straight lines
\begin{equation}
y_i-y_j = 0, \; \mbox{all pairs} \; (i,j) \; i \not= j
\label{4.28}
\end{equation}
cutting the plane into six sectors. They correspond to the six elements of the symmetric 
group $S_3$ (see (\ref{1.11})-(\ref{1.14})). From (\ref{4.20}), (\ref{4.26}) we see that the 
inverse image of a zero trajectory ${\cal C}_n$ consists of six straight lines as well
\begin{equation}
\frac{1}{\sqrt{2}} s^{\frac12} (y_i-y_j) = n\pi, \; n \in \bbbn,
\; \mbox{all pairs} \; (i,j) \; i \not= j
\label{4.29}
\end{equation}
Three of these lines intersect the same sector and these lines intersect each other on the
boundary of the sector, i.e. on the separatrix. In Fig. 2 we present the resulting image
for the identity element of $S_3$ as sector, projected on the $(y_2,y_3)$-plane.
\begin{figure}
\begin{center}
\epsfig{file=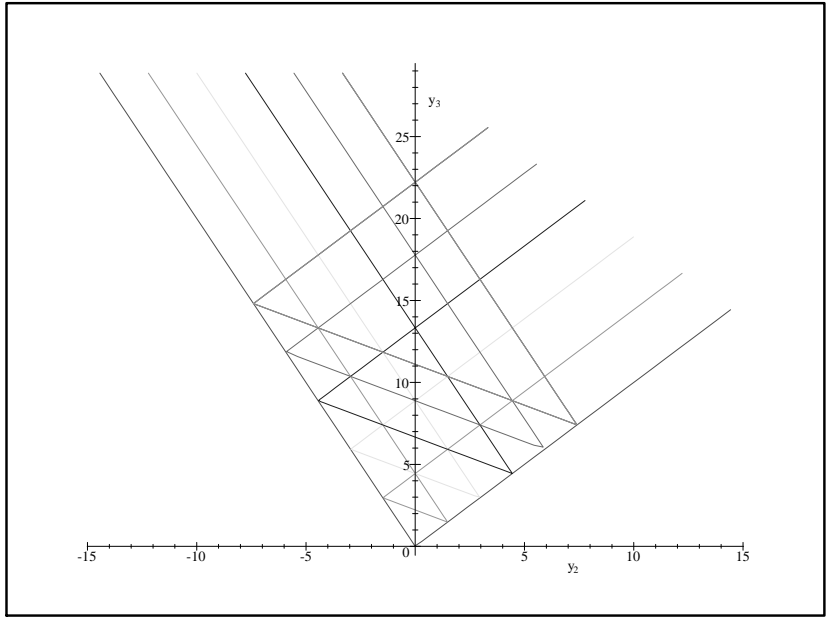,height=12cm,angle=270}
\end{center} 
\label{fig:zeroyy} 
\caption{The sector$E_1$ in $(y_2,y_3)$-plane with zerotrajectories $1 \leq n \leq  5$} 
\end{figure}

\setcounter{equation}{0}
\section{The spectrum}
On the linear space (\ref{1.27})
\begin{equation}
V_n = \bigoplus\limits^n_{m=0} U_m
\label{5.1}
\end{equation}
we consider the quadratic operator
\begin{equation}
-\left( \frac{\partial}{\partial\xi}, \frac{\partial}{\partial\eta} \right) \Gamma(\xi,\eta) \left(
\begin{array}{c} \frac{\partial}{\partial\xi} \\
\frac{\partial}{\partial\eta} \end{array} \right)
\label{5.2}
\end{equation}
(differentiation of $\Gamma$ included) with $\Gamma$ the matrix (\ref{4.5}). This operator
decomposes in two parts ${\cal A} + {\cal B}$ so that for all $m$
\begin{equation}
{\cal A} U_m \subset U_m, \; {\cal B} U_{m+1} \subset U_m
\label{5.3}
\end{equation}
Now we span $U_m$ by monomials of degree $m$
\begin{equation}
\sum^m_{r=0} a^{(m)}_r \xi^r\eta^{m-r}
\label{5.4}
\end{equation}
then the map ${\cal A}$ is
\begin{eqnarray}
a^{(m)}_r \begin{array}{c}
\to \\ {\cal A} \end{array}
a^{\prime (m)}_r &=& [m^2 + \frac73 m + \frac23 r(m-r)] a^{(m)}_r \nonumber \\
& & - \frac23 (m-r+2)(m-r+1) a^{(m)}_{r-2} \nonumber \\
& & - \frac16 (r+2) (r+1) a^{(m)}_{r+2}
\label{5.5}
\end{eqnarray}
\begin{equation}
a^{(m)}_r \begin{array}{c}
\to \\ {\cal B} \end{array} a_r^{\prime(m-1}) = (r+1)(6m-4r-1)a_{r+1}^{(m)}
\label{5.6}
\end{equation}
From (\ref{5.5}), (\ref{5.6}) we recognize that all spaces $U_m$ decompose into direct sums of 
even and odd parts (the power $r$ of $\xi$ is even or odd, respectively)
\begin{equation}
U_m = U_m^{(+)} \oplus U_m^{(-)}
\label{5.7}
\end{equation}
\begin{equation}
U_m^{(+)} = {\rm span} \{ \xi^{2r} \eta^{m-2r} \}^{\left[ \frac m2 \right]}_{r=0}
\label{5.8}
\end{equation}
\begin{equation}
U_m^{(-)} = {\rm span} \{ \xi^{2r+1} \eta^{m-2r-1} \}^{\left[ \frac{m-1}{2} \right]}_{r=0}
\label{5.9}
\end{equation}
so that
\begin{eqnarray}
{\cal A} U_m^{(\pm)} \subset U_m^{(\pm)} \nonumber \\
{\cal B} U_m^{(\pm)} \subset U_{m-1}^{(\mp)}
\label{5.10}
\end{eqnarray}
Correspondingly the set of eigenvalues $\Lambda^{(m)}$ of ${\cal A}$ on $U_m$
decomposes
\begin{equation}
\Lambda_m = \Lambda_m^{(+)} \cup \Lambda_m^{(-)}
\label{5.11}
\end{equation}
The eigenvalues on each subspace $U_m^{(\pm)}$ are non-degenerate for fixed $m$.

For the even part we find 
\begin{eqnarray}
\Lambda_m^{(+)} = \{ \frac13 [4m^2 - (4r-10)m + 4(r-1)^2]; \nonumber \\
r \in \{ 1,2,\ldots, [\frac m2] + 1\} \}
\label{5.12}
\end{eqnarray}
For increasing $r$ the eigenvalues $\lambda_r^{(m)}$ decrease. Moreover we find
\begin{eqnarray}
\Lambda_m^{(-)} &=& \Lambda_m^{(+)}, \quad \mbox{$m$ odd} \nonumber \\
\Lambda_m^{(-)} &=& \Lambda_m^{(+)} - \{m^2 + 2m\}, \quad \mbox{$m$ even}
\label{5.13}
\end{eqnarray}
where $m^2+2m$ is the smallest eigenvalue of $\Lambda_m^{(+)}$.

On $V_n$ (\ref{5.1}) ${\cal A} + {\cal B}$ has the matrix form
\begin{equation}
\left( \begin{array}{cccccccc}
{\cal A}_n, & 0 \\
{\cal B}_n, & {\cal A}_{(n-1)} & 0 &  &  \\
0 & {\cal B}_{(n-1)} & {\cal A}_{(n-2)} & 0 \\
& 0& {\cal B}_{(n-2)} & {\cal A}_{(n-3)} &  \\
&&&&&0 \\
&  & & & & {\cal B}_2 & {\cal A}_1 & 0 \\
& && & & 0 &{\cal B}_1 & {\cal A}_0 
\end{array} \right)
\label{5.14}
\end{equation}
where ${\cal A}_m ({\cal B}_m)$ is the restriction of ${\cal A} ({\cal B})$ to $U_m$.
Assume we want to calculate an eigenvector in $V_n$ which is not in $V_m (m < n)$. If
\begin{equation}
\Lambda_m \cap \Lambda_{m^{\prime}} = \phi \;\, {\rm for} \;\, m \not= m^{\prime}
\label{5.15}
\end{equation}
then
\begin{equation}
{\cal A}_n e_n = \lambda_r^{(n)} e_n, \quad e_n \in U_n 
\label{5.16}
\end{equation}
implies for the component $e_m$ of the same eigenvector in $U_m, \, m < n$
that 
\begin{equation}
e_m = - ({\cal A}_m - \lambda_r^{(m)} E)^{-1} {\cal B}_{(m+1)} e_{(m+1)}
\label{5.17}
\end{equation}
However, (\ref{5.15}) is not correct. Instead we found an infinite sequence of 
counter examples. The smallest ones are
\begin{equation} \begin{array}{rclcl@{\qquad}rclcl}
\lambda_1^{(14)} &=& \lambda_8^{(16)} &=& \frac{868}{3} & \quad
\lambda_1^{(27)} &=& \lambda_{15}^{(31)} &=& 1026 \\
\lambda_2^{(18)} &=& \lambda_9^{(20)} &=& \frac{1336}{3} &
\lambda_6^{(30)} &=& \lambda_{15}^{(32)} &=& \frac{3280}{3} \\
\lambda_3^{(22)} &=& \lambda_{10}^{(24)} &=& 636 &
\lambda_5^{(30)} &=& \lambda_{12}^{(32)} &=& \frac{3364}{3} \\
\lambda_1^{(22)} &=& \lambda_6^{(24)} &=& \frac{2068}{3} & 
\lambda_2^{(30)} &=& \lambda_7^{(32)} &=& \frac{3664}{3} \\
\lambda_4^{(26)} &=& \lambda_{11}^{(28)} &=& \frac{2584}{3} &
\lambda_2^{(31)} &=& \lambda_{16}^{(35)} &=& \frac{3910}{3}
\end{array}
\label{5.18}
\end{equation}
Two eigenvalues can only coincide if $m_1-m_2$ is even and $m_1,m_2$ are large. Consequently
for such coinciding pair with $m_1 < m_2, r_1 < r_2$ the inverse
\[ ({\cal A}_{m_1} - \lambda_{r_2}^{(m_2)} E)^{-1} \]
does not exist. But in addition 
\begin{equation}
{\cal B}_{m_1+1} e_{m_1+1} \not= 0 
\label{5.19}
\end{equation}
since ${\cal B}_m$ (\ref{5.6}) is diagonal and its elements are all non-zero.

Thus the eigenvectors in $V_{m_2}$ belonging to $\lambda_{r_2}^{(m_2)}$ are missing.
Any smooth function in $\xi$ and $\eta$ which can be approximated uniformly over compact
domains by polynomials due to the Weierstra{\ss} approximation theorem cannot be approximated by the
eigenpolynomials of the operator (\ref{5.2}).

This incompleteness can be cured by adding a linear term in the Lie algebra to (\ref{5.2}), e.g.
\begin{equation}
+ \gamma \left( \xi \frac{\partial}{\partial \xi} + \eta \frac{\partial}{\partial \eta} \right), \;
\gamma \notin \bbbq 
\label{5.20}
\end{equation}

In fact we shall see in the subsequent section that such term is essential for self-adjointness of the 
Schr\"odinger operator.

\setcounter{equation}{0}
\section{The Schr\"odinger operator}
We start from (\ref{3.19}) and consider the operator 
\begin{equation}
Q = - \sum_{a,b} \frac{\partial}{\partial \tau_a} g^{-1}_{ab} \frac{\partial}{\partial \tau_b}
+ \gamma \sum_a r_a \frac{\partial}{\partial \tau_a} 
\label{6.1}
\end{equation}
where $\gamma \in \bbbr$ is arbitrary and the functions $r_a(\tau)$ are linear in $\{\tau_a\}$ 
and left otherwise open for a while. The spectrum of $Q$ can be calculated with the method of
Section 5. Then we may have eigenfunctions $\varphi$ and eigenvalues $\epsilon$
\begin{equation}
Q\varphi = \epsilon \varphi 
\label{6.2}
\end{equation}
By an appropriate choice of a gauge function $\chi$
\begin{equation}
\varphi = e^{\chi}\psi 
\label{6.3}
\end{equation}
we want to transform $Q$ into a standard Schr\"odinger operator in $\tau$-space
\begin{equation}
- \Delta \psi + W \psi = \epsilon \psi 
\label{6.4}
\end{equation}
Here $\Delta$ is the Laplace-Beltrami operator 
\begin{equation}
\Delta = \frac{1}{\sqrt{g}} \sum_{a,b} \frac{\partial}{\partial\tau_a} \sqrt{g} g^{-1}_{ab}
\frac{\partial}{\partial\tau_b}
\label{6.5}
\end{equation}
and $W$ is a potential depending on $\chi$ and $\gamma$. The determinant $g$ is computed from
\begin{equation}
g = (\det \{g^{-1}_{ab}\})^{-1} 
\label{6.6}
\end{equation}
and comes out from (\ref{3.19}) as
\begin{eqnarray}
g(s;\tau)^{-1} = - \frac13 \big\{ (4\tau^3_2 + 27\tau^2_3) + 2 s \tau_2(\tau^3_2 + 9\tau^2_3) 
\nonumber \\
+ 2 s^2\tau^2_2\tau^2_3 + \frac12 s^3 \tau^4_3 \big\} 
\label{6.7}
\end{eqnarray}
This function can be expressed by $\{\tau^{\prime}_a\}$ easily as follows from (\ref{4.1})
\begin{equation}
g(s;\tau(\tau^{\prime}))^{-1} = - \frac13 (4\tau^{\prime 3}_2 + 27\tau^{\prime 2}_3) \cdot
(\det \Omega(\tau^{\prime}))^2 
\label{6.8}
\end{equation}
After multiplication with appropriate powers of $s$ we obtain  from (\ref{4.8})
\begin{equation}
\Gamma(\xi,\eta) = - \frac13 D(\xi^{\prime},\eta^{\prime}) (\det H(\xi^{\prime}, \eta^{\prime}
))^2
\label{6.9}
\end{equation}
where $\Gamma$ is now the determinant of the matrix (\ref{4.5}).

In order to fix $\chi$ we start from
\begin{eqnarray}
e^{-\chi} Q e^{\chi} &=& - \sum_{a,b} \left( \frac{\partial}{\partial \tau_a} + \frac{\partial\chi}
{\partial\tau_a}\right) g^{-1}_{ab} \left( \frac{\partial}{\partial\tau_b} + \frac{\partial\chi}
{\partial\tau_b} \right) \nonumber \\
& &+ \gamma \sum_a r_a \left( \frac{\partial}{\partial\tau_a}+\frac{\partial\chi}{\partial\tau_a}
\right)
\label{6.10}
\end{eqnarray}
We match the first order differential operator parts in (\ref{6.4}), (\ref{6.5}) and
(\ref{6.10})
\begin{equation}
-2 \sum_a \frac{\partial\chi}{\partial\tau_a} g^{-1}_{ab} + \gamma r_b = - \sum_a
\frac{\partial}{\partial\tau_a} (\ln \sqrt{g}) g^{-1}_{ab} 
\label{6.11}
\end{equation}
The integrability condition for this differential equation is
\begin{equation}
\frac{\partial}{\partial\tau_a} \sum_c r_c g_{cb} = \frac{\partial}{\partial\tau_b}
\sum_c r_c g_{ca} 
\label{6.12}
\end{equation}
If (\ref{6.12}) is fulfilled we obtain a function $\rho$ so that
\begin{equation}
r_a = \sum_b g^{-1}_{ab} \frac{\partial}{\partial\tau_b} \rho 
\label{6.13}
\end{equation}
In fact, we can show easily that
\begin{equation}
\rho = \ln \sqrt{g}
\label{6.14}
\end{equation}
and
\begin{eqnarray}
r_2 &=& 3 + 2 s\tau_2 \nonumber \\
r_3 &=& 2s\tau_3
\label{6.15}
\end{eqnarray}
fulfill (\ref{6.12}) and (\ref{6.13}). Then (\ref{6.11}) is solved by
\begin{equation}
\chi = \frac12 (1+\gamma) \ln \sqrt{g}
\label{6.16}
\end{equation}
and the potential comes out as
\begin{eqnarray}
W(\tau) &=& - \sum_{a,b} \left\{ \frac{\partial}{\partial\tau_a} \left( g^{-1}_{ab}
\frac{\partial \chi}{\partial\tau_b} \right) + g^{-1}_{ab} \frac{\partial\chi}{\partial\tau_a}
\frac{\partial\chi}{\partial\tau_b} \right\} \nonumber \\ 
& & + \gamma \sum_a r_a \frac{\partial\chi}{\partial\tau_a}
\label{6.17}
\end{eqnarray}
Due to (\ref{6.13}), (\ref{6.15}) the first term in the potential is a constant and we get
\begin{equation}
W(\tau) = \frac14 (\gamma^2-1) \sum_{a,b} g^{-1}_{ab} \frac{\partial \ln \sqrt{g}}{\partial
\tau_a} \frac{\partial \ln \sqrt{g}}{\partial \tau_b} + \, {\rm const}
\label{6.18}
\end{equation}
The wave functions
\begin{equation}
\psi = e^{-\chi} \varphi = g^{- \frac14(1+\gamma)}\varphi
\label{6.19}
\end{equation}
have a zero of order $\frac12(1+\gamma)$ on each trajectory ${\cal C}_n$ and on the separatrix
which can be seen from
\begin{equation}
g^{- \frac12} = {\rm const.} \, |\prod_{i<j} \sin \sqrt{\frac s2} (y_i-y_j)|
\label{6.20}
\end{equation}
which follows from (\ref{4.13}), (\ref{4.20}), (\ref{4.26}) and (\ref{6.9}). Moreover the
spectral degeneracy is lifted if $\gamma$ is irrational (and $s = 1$, say) due to the form
(\ref{6.15}) of $r_2$ and $r_3$ (compare with (\ref{5.20})). Thus we conclude that
\begin{enumerate}
\item[(1)] for $\gamma = 0$ the Schr\"odinger operator is only formally selfadjoint, since
its eigenfunctions form an incomplete set;
\item[(2)] for $\gamma > 0$ the Schr\"odinger operator is self-adjoint if $\gamma \notin \bbbq$
at least. For (see (\ref{1.3}))
\begin{equation}
\gamma = 2 \nu - 1, \quad \nu \in \bbbz
\label{6.21}
\end{equation}
insertion of (\ref{6.20}) into (\ref{6.18}) gives the Sutherland model \cite{7}, which is known
to be exactly soluble.
\end{enumerate}
The wavefunctions $\psi$ have support on any minimal triangle of a sector $E_g$ in $y$-space.
This triangle is bounded by three trajectories
\begin{eqnarray}
{\cal C}_{n_1}, \; {\cal C}_{n_2}, \; {\cal C}_{n_3} : \nonumber \\
n_1 \le n_2 \le n_3 , \; n_3 = n_1 + n_2 \pm 1 
\label{6.22}
\end{eqnarray}

\setcounter{equation}{0}
\section{The $N=4$ Calogero model}
The Calogero model for $N = 3$ is still too special to draw general conclusions from
it, e.g. its Hamiltonian (\ref{1.1}) is separable. In fact we were able to find flat quadratic
Lie algebraic forms also for $N = 4$. With $s, w_3, w_4$ as in (\ref{3.16}), (\ref{3.17}) we make the ansatz
\begin{eqnarray}
g^{-1}_{22} &=& -2\tau_2 - (\tau^2_2 + a_1w_4\tau_4)s \nonumber \\
& & -(a_2w^2_3\tau^2_3 -  a_3w_4\tau_2\tau_4)s^2 \nonumber \\
& & -a_4w^2_4\tau^2_4s^3 
\label{7.1}
\end{eqnarray}
\begin{equation}
g^{-1}_{23} = -3\tau_3 - a_5\tau_2\tau_3s - a_6w_4\tau_3\tau_4s^2
\label{7.2}
\end{equation}
\begin{eqnarray}
g^{-1}_{24} &=& -4 \tau_4 - (a_7\tau_2\tau_4 + a_8w_4^{-1}w^2_3\tau_3^2)s \nonumber \\
& & -a_9w_4\tau^2_4s^2  
\label{7.3}
\end{eqnarray}
\begin{eqnarray}
g^{-1}_{33} &=& - 4w_4w^{-2}_3\tau_4 + w^{-2}_3\tau^2_2 \nonumber \\
& & -(a_{10}w_4w^{-2}_3\tau_2\tau_4 +  a_{11}\tau^2_3)s \nonumber \\
& & -a_{12}w^2_4w^{-2}_3\tau^2_4s^2 
\label{7.4}
\end{eqnarray}
\begin{equation}
g^{-1}_{34} = + \frac12 w^{-1}_4\tau_2\tau_3 - a_{13}\tau_3\tau_4s 
\label{7.5}
\end{equation}
\begin{equation}
g^{-1}_{44} = - 2 w^{-1}_4 \tau_2\tau_4 + \frac34 w^2_3w^{-2}_4\tau^2_3 - a_{14}\tau^2_4s
\label{7.6}
\end{equation}
Here $w_3$ and $w_4$ can be scaled to one by the automorphisms (\ref{3.14}), (\ref{3.15}).
Setting then $s = 0$ the Calogero Riemannian for $N = 4$ appears.

We have to solve the equations resulting from equating all six components of the curvature
tensor to zero. These are polynomial equations and depend on the fourteen variables $\{a_{\alpha}\}$
of (\ref{7.1})--(\ref{7.6}). The number of equations lies somewhere between $10^3$ and $10^4$.
Using the seventy-eight variables $\{C_{\alpha\beta}\}$ from (\ref{2.4}) the number of
equations would have been considerably bigger still. We solved the equations by solving
first all equations with less than thirty terms and inserting then the result into the
bulk of the equations. This trivialized almost all of them. The result is
\begin{eqnarray}
\begin{array}{lcl@{\quad}lcl}
a_1  & = & 3a_7 - 5 & a_8 & = & - \frac38 a_7 + \frac58 \\
a_2  & = & -\frac{3}{16} a^2_7 + \frac58 a_7 - \frac{11}{16} &a_9 & = & 
- \frac12 a_7^2 + \frac32 a_7 - 1 \\
a_3  & = & \frac12 a_7^2 - a_7 + \frac12 & a_{10} & = & a_7 -1 \\
a_4  & = & - \frac14 a_7^3 + a^2_7 - \frac54 a_7 + \frac12 & a_{11} & = & 1 \\
a_5 & = & - \frac14 a_7 + \frac74 & a_{12} & = & - \frac14 a^2_7 + \frac12 a_7 - \frac14 \\
a_6  & = &  \frac18 a^2_7 - \frac12 a_7 + \frac38 & a_{13} & = & \frac14 a_7 + \frac34 \\
a_7  & = & a_7 & a_{14} & = & - a_7 + 3 
\end{array}
\label{7.7}
\end{eqnarray}

Given the Riemannian (\ref{7.1})--(\ref{7.7}) and the unperturbed form 
\begin{eqnarray}
\Gamma^{(0)} & = &  \left ( \begin{array}{ccc} 
                                 -2\xi^{'} & -3\eta^{'}_3 & -4\eta^{'}_4 \\
                                 -3\eta^{'}_3 & -4\eta^{'}_4+\xi^{'2} & \frac12 \xi^{'} \eta^{'}_3 \\
                                 -4\eta^{'}_4  & \frac12 \xi^{'} \eta^{'}_3 & -2\xi^{'}\eta^{'}_4+\frac34\eta^{'2}_3 \\       
                               \end{array} \right ) \label{7.8a}
\end{eqnarray}
we can solve (\ref{4.8}) in terms of three functions 
\begin{eqnarray}
\xi &=& F(\xi^{'},\eta^{'}_3,\eta^{'}_4) \label{7.9a} \\
\eta_{3} &=& G_3(\xi^{'},\eta^{'}_3,\eta^{'}_4) \label{7.10a} \\
\eta_{4} &=& G_4(\xi^{'},\eta^{'}_3,\eta^{'}_4) \label{7.11a} 
\end{eqnarray}
with the Jacobian matrix
\begin{eqnarray}
H & = &  \left ( \begin{array}{ccc} 
                                 \frac{\partial F}{\partial \xi^{'} } & \frac{\partial G_3}{\partial \xi^{'} }  & \frac{\partial G_4}{\partial \xi^{'} }  \\
                                 \frac{\partial F}{\partial \eta^{'}_3 } & \frac{\partial G_3}{\partial \eta^{'}_3 }  & \frac{\partial G_4}{\partial \eta^{'}_3}  \\
                                 \frac{\partial F}{\partial \eta^{'}_4} & \frac{\partial G_3}{\partial \eta^{'}_4}  & \frac{\partial G_4}{\partial \eta^{'}_4}  \\       
                               \end{array} \right ) \label{7.12a}
\end{eqnarray}
We set 
\begin{equation}
w_3=w_4=1 \label{7.13a}
\end{equation}
We find that only $F$ depends on $a_7$ linearly
\begin{equation}
F=F_1+a_7F_2 \label{7.14a}
\end{equation}
and that 
\begin{equation}
F_2=\frac12 G_4 \label{7.15a}
\end{equation}
Consequently det$H$ is independent of $a_7$. The functions $F_1,G_3,G_4$ can be expressed in summed 
up form correspondingly to (\ref{4.14a}),(\ref{4.15a})
\begin{eqnarray}
F_1  =  \frac12 \sum_{\textrm{\scriptsize{perm. of $\{1,\ldots,4\}$}}} & &\sin{ \{ \left (\frac{s}{2} \right)^\frac12 y_i \} }  \sin{ \{ \left (\frac{s}{2} \right)^\frac12 y_j \} }  \nonumber \\ 
&  & \times \cos{ \{ \left (\frac{s}{2} \right)^\frac12 y_k \} }  \cos{ \{ \left (\frac{s}{2} \right)^\frac12 y_l \} }   \label{7.16a}  \\
G_3  =  \frac{2^{\frac12}}{3} \sum_{\textrm{\scriptsize{perm. of $\{1,\ldots,4\}$}}}  & & \sin{ \{ \left (\frac{s}{2} \right)^\frac12 y_i \} }  \sin{ \{ \left (\frac{s}{2} \right)^\frac12 y_j \} }  \nonumber \\ 
&  & \times \sin{ \{ \left (\frac{s}{2} \right)^\frac12 y_k \} }  \cos{ \{ \left (\frac{s}{2} \right)^\frac12 y_l \} }  \label{7.17a} 
\end{eqnarray}
\begin{eqnarray}
G_4 & = & 4 \sin{ \{ \left (\frac{s}{2} \right)^\frac12 y_1 \} } \sin{ \{ \left (\frac{s}{2} \right)^\frac12 y_2 \} }  \nonumber \\ 
& & \times \sin{ \{ \left (\frac{s}{2} \right)^\frac12 y_3 \} } \sin{ \{ \left (\frac{s}{2} \right)^\frac12 y_4 \} } \label{7.18a} 
\end{eqnarray} 
The separatrix
\begin{equation}
D=0 \label{7.19a}
\end{equation}
is given by the Vandermonde determinant 
\begin{equation}
D=-s^6 V(y_1,y_2,y_3,y_4)^2 \label{7.20a}
\end{equation}
or by 
\begin{eqnarray}
D & = & - 4 \det \Gamma^{(0)} \label{7.21a} \\
D & = & 27 \eta_3^4-256\eta_4^3+128 \xi^2\eta_4^2 \nonumber \\ 
  &   &  -16 \xi^4 \eta_4 +4\xi^3\eta_3^2-144\xi\eta_3^2\eta_4 \label{7.22a}
\end{eqnarray}
Moreover we obtain 
\begin{equation}
\det H = \prod_{1\leq i < j \leq4} \frac{\sin{\alpha_{ij}}}{\alpha_{ij}} \label{7.23a}
\end{equation}
with 
\begin{equation}
\alpha_{ij}=\left(\frac{s}{2} \right)^{\frac12} (y_i-y_j) \label{7.24a}
\end{equation}
and 
\begin{eqnarray}
\det \Gamma(\xi,\eta_3,\eta_4) = -\frac14 D(\xi^{'},\eta_3^{'},\eta_4^{'}) \det H(\xi^{'},\eta_3^{'},\eta_4^{'})^2 \label{7.25a}
\end{eqnarray}
In the case $N=4$ the discussion of the Schr\"odinger operator proceeds along the lines of 
section 5. The linear differential operator part in (\ref{6.1}) is now (for $s=w_3=w_4=1$) 
\begin{eqnarray}
r_2 & = & \frac12 (a_7+5)\tau_2 -\frac14 (a_7^2-4a_7+3)\tau_4+6 \label{7.26a} \\
r_3 & = & 3\tau_3 \label{7.27a} \\
r_4 & = & \tau_2-\frac12 (a_7-9)\tau_4 \label{7.28a} 
\end{eqnarray}
The integrability condition (\ref{6.12}) is fulfilled and the solution of the differential 
equation (\ref{6.13}) is (\ref{6.14}) as for $N=3$. The potential formula (\ref{6.18}) remains 
correct and yields the $N=4$ Sutherland model potential. 
\setcounter{equation}{0}
\section{Concluding remarks}
In carrying over these results to general $N$ we must clarify the meaning of the parameter $a_7$ 
in (\ref{7.1})--(\ref{7.7}). Assume 
\begin{eqnarray}
\xi & = & F(\xi^{'},\eta_3^{'}, \ldots,\eta_{N}^{'} ) \nonumber \\
\eta_k & = & G_k(\xi^{'},\eta_3^{'}, \ldots,\eta_{N}^{'} )  \label{8.1a} \\ 
k & \in & \{3,4,\ldots,N\} \nonumber
\end{eqnarray}  
to be a basic diffeomorphism satisfying the equation analogous to (\ref{4.8}). Then 
another basic diffeomorphism is obtained from (\ref{8.1a}) by application of a linear 
map with triangular matrix 
\begin{eqnarray}
\xi & = & F + \sum^{\textrm{\scriptsize{N}}}_{\begin{array}{l}
                        \textrm{\scriptsize{m=4}} \nonumber \\ 
                        \textrm{\scriptsize{m even}} \nonumber \\
                        \end{array}} a_{2m} G_m \\ \label{8.2a}
\eta_k & = & G_k + \sum^{\textrm{\scriptsize{N}}}_{\begin{array}{l}
                        \textrm{\scriptsize{m=k+2}} \\ 
                        \textrm{\scriptsize{m-k even}}  
                        \end{array}} a_{km} G_m  
\end{eqnarray} 
So the flat Riemannian obtained by our algorithm depends on $\frac14(N-2)^2$ ($N$ even)
 respectively $\frac14(N-1)(N-3)$ ($N$ odd) parameters, which drop out of $\det H$ and 
thereby drop out of the Schr\"odinger operator. Despite of the trivial appearance of 
these parameters in the diffeomorphism (\ref{8.2a}), they are connected with an annoying 
complication of the Riemannian as (7.7) shows.

We have not discovered yet new exactly solvable models nor have we been able to transform 
new known integrable models into exactly solvable forms. However, the exactly solvable 
form of the Sutherland models found here differs from the known one \cite{3}. Moreover, 
we have shown that the formalism of the flat quadratic Lie algebraic forms implies an 
algorithm for the deformation of exactly solvable models into other exactly solvable models.   
%
%
%
%
%
%

\end{document}